\newif\ifdraft

\documentclass{sig-alternate-05-2015}
\usepackage{amsmath, amsfonts}
\usepackage[english]{babel}
\usepackage{flushend}
\usepackage{multicol}
\usepackage{color}
\usepackage{algorithm}
\usepackage{algpseudocode}
\usepackage{xspace}
\usepackage{pgfplots}
\usepackage{float}
\newfloat{algorithm}{t}{lop}
\usepackage{etoolbox}
\makeatletter
\patchcmd{\maketitle}{\@copyrightspace}{}{}{}
\makeatother

\title{Adversarial Perturbations Against Deep Neural Networks for Malware Classification}
\numberofauthors{5}
\author{
\alignauthor 
Kathrin Grosse\\
\affaddr{CISPA, Saarland University}\\
\email{\texttt{grosse@cs.uni-saarland.de}}
\alignauthor 
Nicolas Papernot\\
\affaddr{Pennsylvania State University}\\
\email{\texttt{ngp5056@cse.psu.edu}}
\alignauthor
Praveen Manoharan\\
\affaddr{CISPA, Saarland University}\\
\email{\texttt{manoharan@cs.uni-saarland.de}}
\and
\alignauthor 
Michael Backes\\
\affaddr{CISPA, Saarland University and MPI-SWS}\\
\email{\texttt{backes@mpi-sws.org}}
\alignauthor 
Patrick McDaniel\\
\affaddr{Pennsylvania State University}\\
\email{\texttt{mcdaniel@cse.psu.edu}}
}
\date{}

\newcommand{\F}{\ensuremath{\mathbf{F}}\xspace}
\newcommand{\X}{\ensuremath{\mathbf{X}\xspace}}
\newcommand{\J}{\ensuremath{\mathbf{J}\xspace}}
\newcommand{\Y}{\ensuremath{\mathbf{Y}\xspace}}
\newcommand{\classes}{\ensuremath{\mathbf{\mathcal{Y}}}}
\newcommand{\samples}{\ensuremath{\mathbf{\mathcal{X}}}}

\newcommand{\monly}{\texttt{manifest\-only}\xspace}
\newcommand{\onlyf}{\texttt{only\-freq}\xspace}
\newcommand{\onlys}{\texttt{only\-small}\xspace}

\DeclareMathOperator*{\argmax}{\mathsf{arg\,max}}

\begin{document}








\maketitle
\begin{abstract}
Deep neural networks, like many other machine learning models, have recently been shown to lack robustness against adversarially crafted inputs. These inputs are derived from regular inputs by minor yet carefully selected perturbations
that deceive machine learning models into desired misclassifications. Existing work in this emerging  field was largely specific to the domain of image classification,
since the high-entropy of images can be conveniently manipulated without changing the images' overall visual appearance.
Yet, it remains unclear how such attacks translate to more security-sensitive applications such as malware detection--which may pose significant challenges in sample generation and arguably grave consequences for failure.

In this paper, we  show how to construct highly-effective adversarial sample crafting attacks for
neural networks used as malware classifiers.  The application domain of malware classification introduces additional constraints in the adversarial sample crafting problem when compared to the computer vision domain: (i) continuous, differentiable input domains are replaced by discrete, often binary inputs; and (ii) the loose condition of leaving visual appearance unchanged is replaced by requiring equivalent functional behavior. 
We demonstrate the feasibility of these attacks on many different instances of malware classifiers that we trained using the DREBIN Android malware data 
set.
We furthermore evaluate to which extent potential defensive mechanisms against adversarial crafting can be leveraged to the setting of
malware classification. While feature reduction did not prove to have a positive impact, distillation and re-training on adversarially crafted samples show promising results.

\end{abstract}

\section{Introduction\ifdraft ($3$ Columns)\fi}\label{section:introduction}
Deep neural networks transformed the way machine learning solve computational tasks that rely on
high-dimensional data. Examples include dominating Go~\cite{silver2016mastering}, handling auto\-nomous cars~\cite{bojarski2016end} 
and classifying images at a large scale~\cite{krizhevsky2012imagenet}. Neural networks exhibit particularly outstanding results in settings that 
involve large amounts of data. They have also been shown to have the capacity of extracting increasingly complex data representations. Hence it is sound to consider the hypothetical application of neural networks to security-critical domains such as malware classification.

While the benefits of neural networks are undisputed, recent work has shown that like many machine learning models, they lack robustness against adversarially crafted inputs. These inputs are derived 
 from regular inputs by minor yet carefully selected perturbations~\cite{goodfellow2015explaining,papernot2016limitations} that induce models into adversary-desired misclassifications. 
 The vast majority of work in this  emerging field is specific to the domain of image classification.
Images have high entropy and can be conveniently manipulated by changing individual pixels on a real and continuous scale. Changes applied are often hardly visible to our eyes.
The approach that conceptually underlies many recent adversarial machine learning research effort involves gradients of the function $\F$ represented by the learned model (e.g., a neural network, logistic regression, nearest neighbor ...) in order to classify inputs: evaluating it on an input $\X$, one can quickly (and fully automatically) either identify  individual input features that should be perturbed iteratively to achieve misclassification~\cite{papernot2016limitations} or 
compute a suitable minor change for each pixel all at once~\cite{goodfellow2015explaining}. 

Adversarially crafted inputs can hence be used for subverting systems that rely on image classification. An example is decision making for auto\-nomous vehicles, which might pose
security threats in certain situations. However, verifying whether adversarial crafting is also applicable to inherently security-critical domains that differ significantly in terms of input type and set of possible valid perturbations remains largely an open problem. 
Adversarial crafting for malware detection, for example, would arguably entail much more severe consequences. But it also raises challenges
not encountered in the previously well-studied setting of computer vision: inputs have significantly less entropy.  They are usually not represented on a continuous scale of real numbers, but as binary values---an application either uses a certain system call or not. Moreover, approaches perturbing a given application are considered valid only if they do not modify or jeopardize the application functionality. This was a significantly easier task in computer vision settings where this condition was replaced by requiring indistinguishability of images for a human
observer (technically: minimizing the distance between the adversarial original images).

\paragraph{Contributions}
In this paper we show how to successfully perform adversarial crafting attacks on neural networks for malware classification. We employ feed forward neural networks, achieving state-of-the-art performance in detecting malware on the DREBIN data set~\cite{arp2014drebin}. 
To  craft adversarial samples, we follow the method originally proposed by 
Papernot et al.~\cite{papernot2016limitations}, but address challenges that appear in the transition from continuous and differentiable input domains in computer vision to discrete and restricted inputs in malware detection. 

Since to the best of our knowledge there is no mature and publicly available malware detection system that uses neural networks, we develop our own classifier. We train and evaluate it on the DREBIN dataset introduced by Arp et al.~\cite{arp2014drebin}, which contains more than 120,000 android applications samples, among them over 5,000 malware samples. All features were extracted using static analysis on the given application. Our classifier achieves up to $97\%$ accuracy with $7.6\%$ false negatives and $2\%$ false positives, despite minimal effort for hyperparameter selection. This matches state of the art malware detection systems that rely on static features.


To generate the adversarial samples, we adapt an algorithm based on the forward derivative of the attacked neural network, originally presented by Papernot et al.~\cite{papernot2016limitations}. We address additional constraints that appear in malware detection: A) We can only fully add or remove features, but not gradually change them. This contrasts to previous applications of adversarial crafting in computer vision. B) We must preserve the utility of the modified application, which we achieve by only adding features, and only those that do not interfere with the functionality of the application. C) We can only add a restricted amount of features. To simplify matters, we therefore decide to only add entries to the \texttt{AndroidManifest.xml} file. This file is contained in the APK (the \emph{android application package}) of the android application that contains the application's code and that is used to distribute the application to the end user. Despite these restrictions, we achieve up to a $85\%$ misclassification rate on malicious applications. We thus validate that adversarial crafting is indeed viable in security critical domains such as malware detection: neural networks should not be used in such domains without taking precautions for hardening them against adversarial crafting.

As a second contribution, we investigate potential methods for hardening neural network based malware detection systems against adversarial crafting: by applying these mechanisms we 
aim at reducing the sensitivity of networks to adversarial manipulations of their inputs, and thus increase their resilience to adversarial crafting. We first look at the 
impact of feature reduction on the the network's sensitivity to adversarial crafting. In a second step we then consider distillation~\cite{papernot2016distillation} and 
re-training on adversarial samples~\cite{szegedy2013intriguing} which have both been proposed as actual defensive mechanisms in the literature. The findings of our experimental 
evaluation of the aforementioned mechanisms is threefold. First, feature reduction does not protect against adversarial crafting. It can even have adverse effects in that it further 
simplifies crafting adversarial samples. Second, using distillation reduces the misclassification rates, but the range of improvement is rather small. And third, re-training on 
adversarially crafted samples improves the resistance of most neural networks, however the parameter choice for re-training has a noticeable impact.

Our findings show that adversarially crafted inputs pose a threat in security-critical domains where
the behavior of unknown programs is being analyzed and classified. 

\newpage
\section{Background}\label{section:background}
\begin{figure*}[!thp]
	\centering
\includegraphics[width=0.9\linewidth]{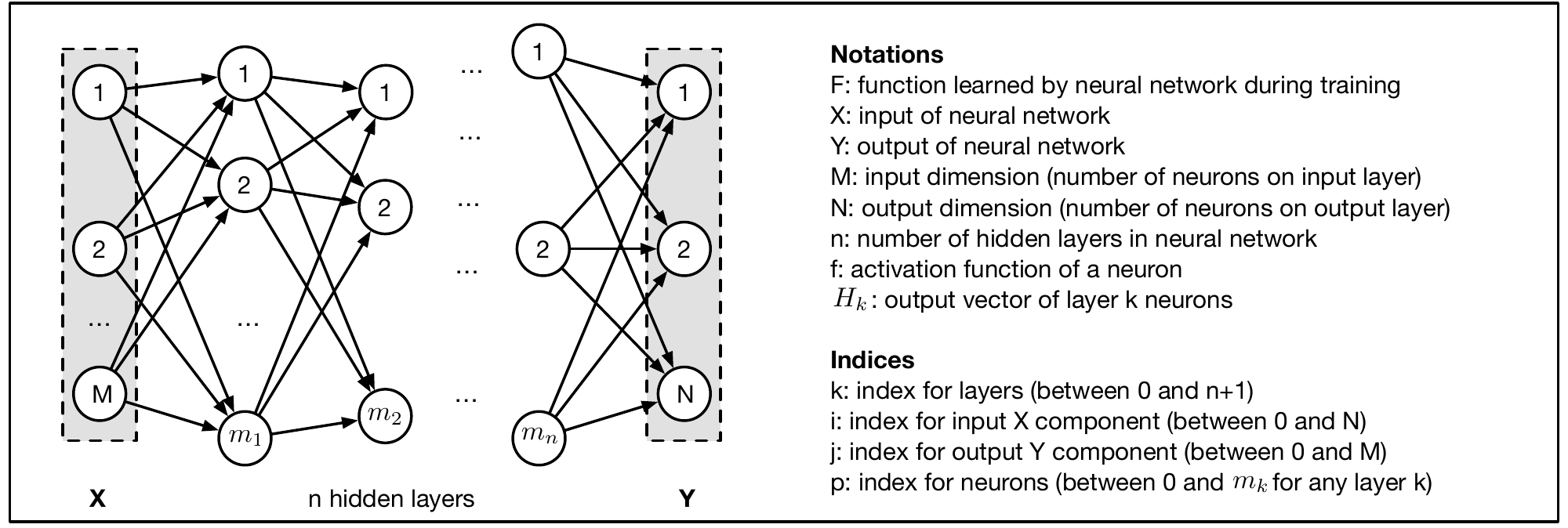}
\caption{The structure of deep feed-forward neural networks, via Papernot et al.~\cite{papernot2016limitations}}\label{fig:neural_network}
\end{figure*}

In this section, we explain the general concepts used in this paper. We first briefly cover background on  neural networks and in some detail how to craft adversarial samples. 

\subsection{Neural Networks}
Neural Networks are machine learning models capable of solving a variety of 
tasks ranging from classification~\cite{krizhevsky2012imagenet, dahl2013large} 
to regression~\cite{neter1996applied} and dimensionality 
reduction~\cite{hinton2006reducing}. They use a graph of elementary computing
units---named \emph{neurons}---organized in \emph{layers} linked to each
other to learn models. Each neuron applies an \emph{activation function},
often non-linear, to its input to produce an output. Figure~\ref{fig:neural_network}, taken from~\cite{papernot2016limitations},
illustrates the general structure of such neural neutworks and also introduces the notation that is used throughout
the paper.

Starting with the model input, each network layer produces an output used as an input by the next layer.
Networks with a single intermediate---\emph{hidden}---layer are qualified as 
\emph{shallow neural networks} whereas models with multiple hidden layers are
\emph{deep neural networks}. Using multiple hidden layers is interpreted as 
hierarchically extracting representations from the input~\cite{Goodfellow-et-al-2016-Book}, eventually producing
a representation relevant to solve the machine learning task and output a 
prediction.

A neural network model $\F$ can be formalized as the composition of
multi-dimensional and parametrized functions $f_i$ each corresponding to a
layer of the network architecture---and a representation of the input:
\begin{equation}
\label{eq:dnn-composition}
\F:\vec{x} \mapsto f_n(...f_2(f_1(\vec{x},\theta_1), \theta_2)..., \theta_n)
\end{equation}
where each vector $\theta_i$ parametrizes layer $i$ of the network $\F$ and includes
weights for the links connecting layer $i$ to layer $i-1$. The set of model 
parameters $\theta=\{\theta_i\}$ is learned during training. For instance, in 
supervised settings, parameter values are fixed by computing prediction 
errors $f(x)-\vec{y}$ on a collection of known input-output 
pairs $(\vec{x},\vec{y})$. 

\subsection{Adversarial Machine Learning}

Neural networks, like numerous machine learning models, have been shown to be 
vulnerable to manipulations of their inputs~\cite{szegedy2013intriguing}. These manipulations
take the form of \emph{adversarial samples}, inputs crafted by adding carefully
selected and often humanly indistinguishable perturbations to inputs so as to
force a targeted model to misclassify the sample. These samples exploit 
imperfections in the training phase as well as the underlying linearity
of components used to learn models---even if the overall model is non-linear like
is the case for deep neural networks~\cite{goodfellow2015explaining}. The space of adversaries was formalized for multi-class deep learning classifiers in a 
taxonomy~\cite{papernot2016limitations}. Adversarial goals can vary from 
simple \emph{misclassification} of the input in a class different from the legitimate 
source class to \emph{source-target misclassification} where samples from
any source class are to be misclassified in a chosen \emph{target class}. 
Adversaries can also be taxonomized by the knowledge of the targeted model they must
possess to perform their attacks. 

In this paper, we study
a binary classifier with only two output classes. Crafting an adversarial 
sample $\vec{x^*}$---misclassified by model $\F$---from a legitimate sample $\vec{x}$ 
can be formalized as the following problem~\cite{szegedy2013intriguing}:
\begin{equation}
\label{eq:adv-sample-crafting-opt-pb}
\vec{x^*}=\vec{x}+\delta_{\vec{x}}=\vec{x}+\min \| \vec{z}\| \ \mathtt{ s.t. }\ \F(\vec{x}+\vec{z}) \neq \F(\vec{x}) 
\end{equation}
where $\delta_{\vec{x}}$ is the minimal perturbation $\vec{z}$ yielding misclassification, 
according to a norm $\|\cdot\|$ appropriate for the input domain. Due to the
non-linearity and non-convexity of models learned by deep neural networks, a closed form solution
to this problem is hard to find. Thus, algorithms were proposed to select perturbations approximatively 
minimizing the optimization problem stated in Equation~\ref{eq:adv-sample-crafting-opt-pb}. The 
\emph{fast gradient sign method} introduced by Goodfellow et al.~\cite{goodfellow2015explaining} linearizes the 
model's cost function around the input to be perturbed and selects a perturbation
by differentiating this cost function with respect to the input itself and not
the network parameters like is traditionally the case during training. 
The \emph{forward derivative} based approach introduced by Papernot et 
al.~\cite{papernot2016limitations} evaluates the model's output sensitivity to
each input component using its Jacobian matrix $J_\F=\left[\frac{\partial \F_j}{\partial x_i}\right]$. A perturbation is then selected 
with adversarial saliency maps, which rank each input component's contribution 
to the adversarial goal by combining components of matrix $J_\F$. Both the fast 
gradient sign and forward derivative methods require full knowledge of the 
targeted model's architecture and parameters. However, a black-box attack 
leveraging both of these approaches to target unknown remotely hosted 
deep neural networks was proposed in~\cite{papernot2016practical}. It first approximates the targeted model by 
querying it for output labels to train a substitute model, which is then used to
craft adversarial samples also misclassified by the originally targeted model.

Machine learning models deployed in adversarial settings therefore need to 
be robust to manipulations of their inputs~\cite{mpc16}. Solutions were proposed
to improve the training algorithms used for deep neural networks, yielding models 
more robust to such perturbations. Goodfellow et al. 
demonstrated that explicitly training with adversarial samples reduced the
error rate of models on samples crafted against the resulting improved 
model~\cite{goodfellow2015explaining}. Papernot et al. proposed the use of 
distillation---training with class probabilities produced by a teacher model instead
of labels---as a defense mechanism, and showed that this effectively reduces the sensitivity of models to
small perturbations~\cite{papernot2016distillation}. Warde-Farley et 
al.~\cite{WardeFarley16} evaluated a simplified
variant of distillation training models on softened indicator vectors instead of
probabilities. They showed error rates reductions on samples crafted using 
the fast gradient sign method.  These solutions however do not completely prevent misclassification of
adversarial samples, which thus remains an open-problem.

\section{Methodology}\label{section:methodology}
This section describes the approach to adversarial crafting for malware detection.
We start by describing how we represent applications for the classification, and how we train the classifiers, detailing which configurations we choose for the neural networks. Thereafter, we describe in detail how we craft adversarial samples based on the forward derivative of the trained neural network, and detail the restrictions on crafting adversarial samples for malware detection (we only add features to ensure functionality is  preserved).



\subsection{Application Model}\label{section:app_model}
Before we can start training a neural network based malware detection system, we first have to decide on a representation of applications that we use as input to our classifier. In this work, we focus on statically determined
features of applications. As a feature, we understand some property that the statically evaluated code of the application exhibits. This includes for instance whether the application uses a specific system call or not, as well as  a usage of specific hardware components or  access to the Internet.

A natural way to represent such features is using \emph{binary indicator vectors}: Given features $1,\ldots,M$, we represent an 
application using the binary vector $\X\in\{0,1\}^M$, where $X_i$ indicate whether the application exhibits feature $i$, i.e. $\X_i=1$, or not, i.e. $\X_i=0$.  Due to the varied nature of applications that are available, $M$ will typically be very large, while each  single application only exhibits very few features. 

This leads to very sparse feature vectors, and overall, a very sparsely populated space of applications in which we try to successfully separate malicious from benign applications. Neural networks have shown to be very successful at separating classes in sparse populated domains. Hence, we will use them to build our malware detection system. 
\subsection{Training the Malware Classifier}\label{section:training_classifier}
To the best of our knowledge, there is no publicly available malware detection system based on neural 
networks that considers static features. While Dahl et al.~\cite{DBLP:conf/icassp/DahlSDY13} use a neural 
networks to classify malware, their approach uses random projections and considers dynamic data. Since perturbing dynamically gathered features is a lot more challenging than modifying static features, we stick to the simpler, static case in this work and leave the dynamic case for future work.

Convolutional neural networks are common architectures for computer vision tasks, and recurrent neural networks for natural language processing and hand-writing recognition. These architectures, however, take advantage of special properties of their input domains to improve their classification performance. On the one hand, convolutional neural networks work well on input containing translation invariant properties, which can be found in images~\cite{mallat2016understanding}. Recurrent neural networks, on the other hand, work well with input data that needs to be processed sequentially~\cite{graves2013generating}.

The binary indicator vector $\X$ we use to represent an application does not possess any of the above structural properties. We therefore stick to regular, feed-forward neural networks as described in Section~\ref{section:background} to solve our malware classification task. Regular feed-forward neural networks are known to not work as well on established use cases as the structured networks mentioned above. However the absence of such structures in our input domain only leaves unstructured feed-forward neural networks. As we will see in Section~\ref{section:evaluation}, these work well enough for our use case.

We train several classifiers using varying configurations: while each network takes the same binary indicator vector $\X$ as input, they differ in the amount of hidden layers (between one and four). Furthermore, we also vary the amount of neurons per layer, ranging between 10 and 300.

We use the rectified non-linearity as the activation function for each hidden neuron in our network, i.e.
\[\forall p\in[1,m_k]: f_{k,p}(x) = \max(0,x)\]
with
\[x=\sum_{j=1}^{m_{k-1}} w_{j,p}\cdot x_j + b_{j,p},\]
where the weight $w_{j,p}$ and the bias $b_{j,p}$ are trained values. After the final hidden layer, we use a softmax layer to normalize the output of the network to a probability distribution, i.e. the output of the network is computed by
\begin{equation}
\F_i(\X) = \frac{e^{x_i}}{e^{x_0}+e^{x_1}}~,~ x_i = \sum_{j=1}^{m_{n}}w_{j,i}\cdot x_j+b_{j,i}
\label{eq:softmax}
\end{equation}
To train our network, we use standard gradient descent with batches of size $1000$ that are split into training and validation sets, using $10$ training epochs per iteration. The performance of the thus trained network is hugely influenced by the choice of the gradient descent \emph{hyperparameters}: these parameters are not trained, but set at the beginning to control the behavior of the gradient descent algorithm. Usually, a large of effort is put toward finding the ideal hyperparameters for a given use case. In our case, we choose these hyperparameters based on previous experience and do not perform an exhaustive search for ideal hyperparameters. For instance, we choose a dropout of $50\%$ between each hidden layer avoid over-fitting, i.e. the output of $50\%$  of all neurons in each layer is ignored and set to 0. As we will see in the evaluation, we still achieve acceptable performance results. However, we expect that the classification performance could be greatly increased by putting more effort into the selection of these hyperparameters. 

Since the DREBIN dataset that we use for our evaluations (cf. Section~\ref{section:evaluation}) has a fairly unbalanced ratio between malware and benign applications, we experiment with different ratios of malware in each training batch to compare the achieved performance values. The number of training iterations is then set in such a way that all malware samples are at least used once. We evaluate the classification performance of each of these networks using accuracy, false negative and false positive rates as performance measures. Afterwards, we evaluate the best performing networks against the adversarial crafting attack we discuss next.

\subsection{Crafting Adversarial Malware Samples}\label{section:crafting_samples}
\begin{algorithm}[t]
\caption{\textbf{Crafting Adversarial Samples for Malware Detection}}\label{sampling_algorithm}
\begin{algorithmic}[1]
\Require $\mathbf{x}$, $\mathbf{y}$, $\mathbf{F}$, $\mathbf{k}$, $\mathbf{I}$
\State $\mathbf{x}^*\leftarrow \mathbf{x}$
\State $\Gamma=\{1\dots|\mathbf{x}|\}$
\While{$\argmax_j \mathbf{F_j}(\mathbf{x}^*)\neq \mathbf{y}$ and $||\delta_\mathbf{X}||<\mathbf{k}$}
	\State Compute forward derivative $\nabla \mathbf{F}(\mathbf{x}^*)$
	\State $i_{max} = \argmax_{j\in\Gamma\cap\mathbf{I},X_j=0} \frac{\partial \mathbf{F}_y(\mathbf{X})}{\partial \mathbf{X}_j}$
	\If{$i_{max} \leq 0$} \State \Return Failure \EndIf
	\State $\mathbf{x}^*_{i_{max}}=1$
	\State $\delta_\mathbf{x} \leftarrow \mathbf{x}^*-\mathbf{x}$
\EndWhile
\State \Return $\mathbf{x}^*$
\end{algorithmic}
\end{algorithm}
The goal of adversarial sample crafting in malware detection is to mislead the detection system, causing the classification for a particular application to change according to the attackers wishes. 

Describing more formally, we start with $X \in\{0,1\}^m$, a binary indicator vector that indicates which features are present in an application. Given $X$, the classifier $\F$ returns a two dimensional vector 
$\F(\X)= \left[\F_0(\X),\F_1(\X)\right]$ with $\F_0(\X)+\F_1(\X) = 1$ that encodes the classifiers belief that $\X$ is either benign ($\F_0(\X)$) or malicious ($\F_1(\X)$). We take as the classification result $y$ the option that has the higher probability, i.e. $y = \argmax_i \F_i(\X)$.
The goal of adversarial sample crafting now is to find a small perturbation $\delta$ such that the classification results $y'$ of $\F(X+\delta)$ is different from the original results, i.e. $y'\neq y$. We denote $y'$ as our \emph{target class} in the adversarial crafting process.

Our goal is to have a malicious application classified as benign, i.e. given a malicious input $\X$, we want a classification results $y'=0$. The opposite case is to misclassify a benign application as malicious. While this is also possible, we assume that the perturbation of the application will be performed by the original author of this application. Since an honest author has no interest in having his benign application classified as malware, we ignore this case. 

We adopt the adversarial crafting algorithm based on the jacobian matrix
\[\J_\F = \frac{\partial \F(\X)}{\partial \X} = \left[ \frac{\partial \F_i(\X)}{\partial \X_j}\right]_{i\in{0,1}, j\in[1,m]}\] 
of the neural network $\F$ put forward by Papernot et al.~\cite{papernot2016limitations} and which we already discussed in Section~\ref{section:background}. Despite it originally being defined for images, we show that an adaptation to a different domain is fairly straight forward.

To craft an adversarial sample, we take mainly two steps. In the first, we compute the gradient of $\F$ with respect to $\X$ to estimate the direction in which a perturbation in $\X$ would change $\F$'s output. In the second step, we choose a perturbation $\delta$ of $\X$ with maximal positive gradient into our target class $y'$. For malware misclassification, this means that we choose the index 
$i = \argmax_{j\in[1,m], \X_j=0} \F_0(\X_j)$ that maximizes the change into our target class $0$ by changing $\X_i$. Note that we only consider positive changes for positions $j$ at which $\X_j=0$, which correspond to adding features the application represented by $\X$ 
(since $\X$ is a binary indicator vector). We discuss this choice in  Section~\ref{section:restrictions}.

Ideally, we keep this change small to make sure that we do not cause a negative change of $\F$ due to intermediate changes of the gradient. For computer vision, this is not an issue since the values of pixels are continuous and can be changes in very small steps. In the malware detection case, however, we do not have continuous data, but rather discrete input values: 
since $\X\in{0,1}^m$ is a binary indicator vector, our only option is to increase one component in $\X$ by exactly $1$ to retain a valid input to $\F$.

The adversarial sample crafting process is iterative: after computing the gradient, we choose a feature whose   gradient is the largest for our target class and change it's value in $\X$ (i.e. by making corresponding changes to the application) to obtain our new input vector $\X^{(1)}$. We then recompute the gradient under this new input $\X^{(1)}$ and find the second feature to change. We repeat this process until either a) we reached the limit for maximum amount of allowed changes or b) we successfully cause a misclassification.  A pseudo-code implementation of the algorithm is given in Algorithm~\ref{sampling_algorithm}. It is largely  similar to the algorithms presented in~\cite{papernot2016limitations}, with the difference in the discrete changes of the input vector due to the input domain and also the additional restrictions below.

\subsection{Restrictions on Adversarial Crafting}\label{section:restrictions}
To make sure that modifications caused by the above algorithms do not change the application too much, we bound 
the maximum distortion $\delta$ applied to the original sample. As in the computer vision case, we only allow 
distortions $\delta$ with $\lVert \delta \rVert \leq k$. We differ, however, in the norm that we apply: in computer vision, the $L_\infty$ norm is often used to bound the maximum change. In our case, each modification to an entry will always change its value by exactly $1$, therefore making the $L_\infty$ norm inappropriate. We instead use the $L_1$ norm to bound the overall number of features modified.   

While the main goal of adversarial crafting is to achieve misclassification, for malware detection, this cannot happen at the 
cost of the application's functionality: feature changes determined by Algorithm~\ref{sampling_algorithm} can cause the 
application in question to lose its functionality in parts or completely. To avoid this case, we adopt the following 
additional restrictions on the adversarial crafting algorithm: we only add features, and only add those that do not 
interfere with other features already present in the application. This protects us from unknowingly destroying the 
applications functionality. Formally, we encode the above restriction through the index set $I$: it contains the 
indices corresponding to features that can be added without affecting the applications functionality.   

In Section~\ref{section:evaluation}, we show that we can successfully craft adversarial samples despite these additional restrictions.

\begin{table}[t]
\centering
\begin{tabular}{l|c|c|c|c}
ID & Name & Manifest & Code & \# \\ \hline
$S_1$ & Hardware Components	& \checkmark	& 			& 4513\\
$S_2$ & Permissions 		& \checkmark	& 			& 3812 \\
$S_3$ & Components 			& \checkmark	& 			& 218951\\
$S_4$ & Intents 			& \checkmark	& 			& 6379\\
$S_5$ & Restr. API Calls	& 				& \checkmark& 733\\	
$S_6$ & Used Permissions	& 				& \checkmark& 70\\
$S_7$ & Susp. API Calls		& 				& \checkmark& 315\\	
$S_8$ & Network Addresses	& 				& \checkmark& 310447
\end{tabular}
\caption{Feature Types, where they are collected and their cardinality.}\label{table:feature_types}
\end{table}

\section{Experimental Evaluation\ifdraft (6-8 Columns)\fi}\label{section:evaluation}
We evaluate the training of the neural network based malware detector and adversarial sample-induced misclassification of inputs on it. Through our evaluation, we want to validate the following two hypotheses: first, that the neural network based malware classifier achieves performance comparable to state of the art malware classifiers (on static features) presented in the literature. Second, the adversarial sample crafting algorithm discussed in Section~\ref{section:crafting_samples} allows us to successfully mislead the neural network we trained. As a measure of success, we consider the misclassification rate achieved by the adversarial crafting algorithm. The misclassification rate is defined by the percentage of malware samples that are misclassified after applying the adversarial crafting algorithm, but were correctly classified before that. 

\subsection{Dataset}
We base our evaluations on the DREBIN dataset, originally introduced by Arp et al.~\cite{arp2014drebin}: DREBIN contains 129.013 android applications, of which 123,453 are benign and 5,560 are malicious. Extracted static features are provided for all applications. In total, the dataset contains 545,333 features that are divided into $8$ feature classes, each of which is represented by a binary value that indicates whether the feature is present in an application or not. This directly translates to the binary indicator 
vector $\X\in\{0,1\}^M$ used to represent applications, with $M=545,333$. 

The $8$ feature classes in DREBIN cover various aspects of android applications, including: A) Permissions and hardware component access requested by each application (e.g. for \texttt{CAMERA} or \texttt{INTERNET} access). B) Restricted and suspicious (i.e. accessing sensitive data, e.g. \texttt{getDeviceID()}) API-calls made by the applications. C) application components such activities, service, content provider and broadcast receivers used by each applications, and D) intents used by applications to communicate with other applications. Table~\ref{table:feature_types} lists each feature class and its cardinality.

In Table~\ref{table:feature_numbers} we give average and quantile statistics on the amount of features exhibited by the applications in
DREBIN. Given these numbers, we decide to set our distortion bound $k = 20$ -- assuming we are modifying an application of average size, it still remains within the two main quartiles when adding at most 20 features.
\begin{table}[t]
\centering
\begin{tabular}{l |c|c|c|c}
 & 1st Quantile & Mean & 3rd Quantile & max \\ \hline
all &  23	&  48	&  61 & 9661\\
malicious & 35 		& 62	& 83 & 666
\end{tabular}
\caption{Some basic statistics on the number of features per app in the DREBIN data set. }\label{table:feature_numbers}
\end{table}  

Since the DREBIN data set contains Android applications, we decide to only add features that can be added through 
modifications in the \texttt{AndroidManifest.xml} file of the android application's APK. The manifest is used by the application to announces its components (i.e. its activities, services, broadcast receivers, and content providers),
the permissions it requests and further information about the application the system needs to run the application. Changes in the manifest are particularly easy to implement, since they only incur an additional line in the manifest and do not cause any interference with the application's functionality. Changes to the code, on the other hand, would require more effort and would have to be handled more carefully. Table~\ref{table:feature_types} lists where each feature class in DREBIN originates from, identifying those features that originate in the manifest and that we will consider in the adversarial crafting algorithm. In Algorithm~\ref{sampling_algorithm}, we represent the set of valid features for modification by the index set $I$.


\subsection{Malware Detection}
\begin{table}[t]
\scriptsize
\centering
\begin{tabular}{c|c| c |c|c|c|c}
Classifier & MWR & Accuracy 	& FNR 		& FPR 	\\ \hline
Arp et al.~\cite{arp2014drebin} & $-$	& $-$ 		& $6.1$& $1$ \\
Sayfullina et al.~\cite{DBLP:conf/trustcom/SayfullinaEKPML15} & $-$ & $-$ & $0.1$ & $17.9$ \\
$[200]$ 	 & 0.4 	 & 97.83 	 & 8.06 	 & 1.86 \\
$[200]$ 	 & 0.5 	 & 95.85 	 & 5.41 	 & 4.06 \\
$[10, 10]$ 	 & 0.3 	 & 97.59 	 & 16.37 	 & 1.74 \\
$[10, 10]$ 	 & 0.4 	 & 94.85 	 & 9.68 	 & 4.90 \\
$[10, 10]$ 	 & 0.5 	 & 94.75 	 & 7.34 	 & 5.11 \\
$[10, 200]$ 	 & 0.3 	 & 97.53 	 & 11.21 	 & 2.04 \\
$[10, 200]$ 	 & 0.4 	 & 96.14 	 & 8.67 	 & 3.6 \\
$[10, 200]$ 	 & 0.5 	 & 94.26 	 & 5.72 	 & 5.71 \\
$[200, 10]$ 	 & 0.3 	 & 95.63 	 & 15.25 	 & 3.86 \\
$[200, 10]$ 	 & 0.4 	 & 93.95 	 & 10.81 	 & 5.82 \\
$[200, 10]$ 	 & 0.5 	 & 92.97 	 & 8.96 	 & 6.92 \\
$[50, 50]$ 	 & 0.3 	 & 96.57 	 & 12.57 	 & 2.98 \\
$[50, 50]$ 	 & 0.4 	 & 96.79 	 & 13.08 	 & 2.73 \\
$[50, 50]$ 	 & 0.5 	 & 93.82 	 & 6.76 	 & 6.11 \\
$[50, 200]$ 	 & 0.3 	 & 97.58 	 & 17.30 	 & 1.71 \\
$[50, 200]$ 	 & 0.4 	 & 97.35 	 & 10.14 	 & 2.29 \\
$[50, 200]$ 	 & 0.5 	 & 95.65 	 & 6.01 	 & 4.25 \\
$[200, 50]$ 	 & 0.3 	 & 96.89 	 & 6.37 	 & 2.94 \\
$[200, 50]$ 	 & 0.4 	 & 95.87 	 & 5.36 	 & 4.06 \\
$[200, 50]$ 	 & 0.5 	 & 93.93 	 & 4.55 	 & 6.12 \\
$[100, 200]$ 	 & 0.4 	 & 97.43 	 & 8.35 	 & 2.27 \\
$[200, 100]$ 	 & 0.4 	 & 97.32 	 & 9.23 	 & 2.35 \\
$[200, 100]$ 	 & 0.5 	 & 96.35 	 & 6.66 	 & 3.48 \\
$[200, 200]$ 	 & 0.1 	 & 98.92 	 & 17.18 	 & 0.32 \\
$[200, 200]$ 	 & 0.2 	 & 98.38 	 & 8.74 	 & 1.29 \\
$\mathbf{[200, 200]}$ 	 & \textbf{0.3} 	 & \textbf{98.35} 	 & \textbf{9.73} 	 & \textbf{1.29} \\
$\mathbf{[200, 200]}$ 	 & \textbf{0.4} 	 & \textbf{96.6} 	 & \textbf{8.13} 	 & \textbf{3.19} \\
$\mathbf{[200, 200]}$ 	 & \textbf{0.5} 	 & \textbf{95.93} 	 & \textbf{6.37} 	 & \textbf{3.96} \\
$[200, 300]$ 	 & 0.3 	 & 98.35 	 & 9.59 	 & 1.25 \\
$[200, 300]$ 	 & 0.4 	 & 97.62 	 & 8.74 	 & 2.05 \\
$[300, 200]$ 	 & 0.2 	 & 98.13 	 & 9.34 	 & 1.5 \\
$[300, 200]$ 	 & 0.4 	 & 97.29 	 & 8.06 	 & 2.43 \\
$[200, 200, 200]$ 	 & 0.1 	 & 98.91 	 & 17.48 	 & 0.31 \\
$[200, 200, 200]$ 	 & 0.4 	 & 97.69 	 & 10.34 	 & 1.91 \\
$[200, 200, 200, 200]$ 	 & 0.4 	 & 97.42 	 & 13.08 	 & 2.07 \\
$[200, 200, 200, 200]$ 	 & 0.5 	 & 97.5 	 & 12.37 	 & 2.01 \\
\end{tabular}
\caption{Performance of the classifiers. Given are used malware ratio (MWR), accuracy, false negative rate (FNR) and false positive rate (FPR).}
\label{table:classifier}
\end{table}
We train numerous neural network architecture variants, according to the training procedure described in Section~\ref{section:methodology}. Our baseline architecture includes 2 hidden layers of 200 neurons each. From here, we vary the number of neurons per layer (from 200 to 10, 50, 100 and 300), and the number of layers (from 2 to 1, 3 and 4). We also vary the malware ratio in each training batch by steps of 0.1 from 0.1 to 0.5 and measure its impact on the overall performance of the neural network in correctly classifying the applications in the DREBIN dataset.

The results for the different neural networks can be found in Table~\ref{table:classifier}. WE first list previous classifiers from the literature, then the architecture (in neurons per layer) that we trained. 
As performance measures we consider the overall accuracy on the DREBIN dataset, as well as the false positive and false negative rates. 

Using the malware ratio in our training batches as a parameter to be optimized, we achieve false negative rates at a level comparable to state of the art classifiers. This, however, happens at a trade-off with overall accuracy and false positive rates. 
In comparison, Arp et al.~\cite{arp2014drebin} achieve a $6.1\%$ false negative rate at a $1\%$ false positive rate. Sayfullina et al.~\cite{DBLP:conf/trustcom/SayfullinaEKPML15} even achieve a $0.1\%$ false negative rate, however at the cost of 
$17.9\%$ false positives. 

We can observe high accuracy ($> 90\%$) across results. The network architecture has some impact on the 
trade-off between accuracy, false positive, and false negative rates at the various malware ratios. However, no 
clear trends can be observed that would indicate how many neurons should be chosen on the first and second layer. 
Overall, the baseline architecture with 200 neurons on 2 layers each achieves, according to our experimental setup, 
the best trade-off between false positive, false negatives and overall accuracy. With this 
architecture, we achieve around $98\%$ overall accuracy, with about $7\%$ false negatives and $3.3\%$ false positives. As discussed in Section~\ref{section:training_classifier}, we expect that this performance can further greatly be improved by searching for hyperparameters better fitting this use case. Overall, we thus validate our hypothesis that a neural network based classifier can successfully classify the DREBIN malware data set.

\subsection{Adversarial Malware Crafting}
\begin{table}[t]
\centering
\scriptsize
\begin{tabular}{c|c|c|c}
Classifier & MWR & MR & Distortion 	\\ \hline
$[200]$ 	& 0.4 	& 81.89 	& 11.52 \\
$[200]$ 	& 0.5 	& 79.37 	& 11.92 \\
$[10, 10]$ 	& 0.3 	& 69.62 	& 13.15 \\
$[10, 10]$ 	& 0.4 	& 55.88 	& 16.12 \\
$[10, 10]$ 	& 0.5 	& 84.05 	& 11.48 \\
$[10, 200]$ 	& 0.3 	& 75.47 	& 12.89 \\
$[10, 200]$ 	& 0.4 	& 55.70 	& 14.84 \\
$[10, 200]$ 	& 0.5 	& 57.19 	& 14.96 \\
$[200, 10]$ 	& 0.3 	& 50.07 	& 14.96 \\
$[200, 10]$ 	& 0.4 	& 35.31 	& 17.79 \\
$[200, 10]$ 	& 0.5 	& 36.62 	& 17.49 \\
$[100, 200]$ 	& 0.4 	& 74.93 	& 12.87 \\
$[200, 100]$ 	& 0.4 	& 71.42 	& 13.12 \\
$[200, 100]$ 	& 0.5 	& 73.02 	& 12.98 \\
$[50, 50]$ 	& 0.3 	& 61.71 	& 15.37 \\
$[50, 50]$ 	& 0.4 	& 60.02 	& 14.7 \\
$[50, 50]$ 	& 0.5 	& 40.97 	& 17.64 \\
$[50, 200]$ 	& 0.3 	& 79.25 	& 11.61 \\
$[50, 200]$ 	& 0.4 	& 69.44 	& 13.95 \\
$[50, 200]$ 	& 0.5 	& 64.66 	& 15.16 \\
$[200, 50]$ 	& 0.3 	& 66.55 	& 14.99 \\
$[200, 50]$ 	& 0.4 	& 58.31 	& 15.76 \\
$[200, 50]$ 	& 0.5 	& 62.34 	& 14.54 \\
$[200, 200]$ 	& 0.1 	& 78.28 	& 10.99 \\
$[200, 200]$ 	& 0.2 	& 63.49 	& 13.43 \\
$\mathbf{[200, 200]}$ 	& \textbf{0.3} 	& \textbf{63.08} 	& \textbf{14.52} \\
$\mathbf{[200, 200]}$ 	& \textbf{0.4} 	& \textbf{64.01} 	& \textbf{14.84} \\
$\mathbf{[200, 200]}$ 	& \textbf{0.5} 	& \textbf{69.35} 	& \textbf{13.47} \\
$[200, 300]$ 	& 0.3 	& 70.99 	& 13.24 \\
$[200, 300]$ 	& 0.4 	& 61.91 	& 14.19 \\
$[300, 200]$ 	& 0.2 	& 69.96 	& 13.62 \\
$[300, 200]$ 	& 0.4 	& 63.51 	& 14.01 \\
$[200, 200, 200]$ 	& 0.1 	& 75.41 	& 10.50 \\
$[200, 200, 200]$ 	& 0.4 	& 71.31 	& 13.08 \\
$[200, 200, 200, 200]$ 	& 0.4 	& 62.66 	& 14.64 \\
\end{tabular}
\caption{Performance of our adversarial sampling strategy. The misclassification rates (MR) and required average distortion (in number of added features) with a threshold of 20 modifications are given for various architectures and malware ratio (MWR).}\label{table:misclassification}
\end{table}

We next apply the adversarial crafting algorithm described in Section~\ref{section:methodology} and observe on how many occasions we can successfully mislead our neural network based classifiers. We quantify the performance of our algorithm through the achieved misclassification rate, which measures the amount of previously correctly classified malware that is misclassified after the adversarial crafting. In addition, we also measure the average number of modifications required to achieve the measured misclassification rate to assess which architecture provided a harder time being mislead. As discussed above, allow at most $20$ modification to any of the malware applications.

The performance results are listed in Table~\ref{table:misclassification}. As we can see, we achieve misclassification rates from at least $50\%$ in the case of the two layer neural network with 200 and 10 neurons in each layer, to up to $84\%$ in the case of the two layer network with 10 neurons in both layers. Again, we cannot directly observe any rule that directly connects network architecture to resistance against adversarial crafting. However, we can observe that the malware ratio used in the training batches is correlated to the misclassification rate: a higher malware ratio in general results in a lower misclassification rate, albeit with exceptions (e.g. in the case of the 2 layer network with 10 neurons in each).

Still, due to the fact that our algorithm is able to successfully mislead most networks for a large majority of malware samples, we validate the hypothesis that our adversarial crafting algorithm for malware can be used to mislead neural network based malware detection systems.  
 
\subsection{Discussion}
As our evaluation results show, adversarial sampling is a real threat also for neural network based malware detection. We therefore confirm that findings from applications in the domain of computer vision can successfully be transferred to more security
critical domains such as malware detection. Despite heavy restrictions on the type of modifications we are allowed to undertake, we are able to mislead our classifiers on about $60\%$ to $80\%$ of the malicious application samples (depending on considered architecture and ignoring corner cases). While this does not match the misclassification rates of up to $97\%$ for misclassifying images of digits reported by Papernot et al.~\cite{papernot2016limitations}, a reduction in misclassification performance was to be expected given the challenge inherent to the malware detection domain.

In the next part of this paper, we are going to consider possible defensive mechanisms to harden our neural network against adversarial crafting. To ensure comparability, we will restrict ourselves to the $200\times200$ neurons architecture for our neural network models (i.e. 2 layers with 200 neurons each). It is trained with malware ratios between $0.3$ and $0.5$ for our subsequent evaluations, and we will use the 
networks of the same architecture and with the same malwario rates as our comparison baselines. In the above evaluations, this configuration provides a reasonable trade-off between false positive and false negative rates while retaining an overall high accuracy. Furthermore, this architecture remains comparatively (outside of corner cases) resistant to adversarial crafting, only being misled in about $64\%$ of the cases while at the same time requiring a comparatively large amount of modifications.
\section{Defenses\ifdraft (2-3 Columns)\fi}\label{section:defenses}
In this section, we investigate potential defensive mechanisms for hardening our neural network-based malware detector against 
adversarial crafting. We measure the effectiveness of each these mechanisms by determining the reduction in misclassification rates, i.e. we determine the difference between the misclassification rates
achieved against regular networks, and those to which we applied the mechanisms. Again, as in the previous section, the misclassification rate is defined as the percentage of malware samples that are misclassified after the application of the adversarial crafting algorithm, but were correctly classified before. 

We first take a look at feature reduction as a potential defensive mechanism: by reducing the number of features we consider for classification, we hope to reduce the neural network's sensitivity to changes in the input, as well as reduce the number of feature options the adversary has for distorting an application and causing a misclassification. 
We consider simple feature reduction, a naive way to include features based on where and how often they appear, as well as a more involved feature reduction based on the mutual information. Our goal is to evaluate whether feature reduction in general can be used to harden a neural network against adversarial crafting. 

After looking at feature reduction, we consider two defensive mechanisms already proposed in the literature: distillation, as introduced by Papernot et al.~\cite{papernot2016distillation}, and re-training on adversarial samples, following methodology described by Szegedy et al.~\cite{szegedy2013intriguing}.
\subsection{Simple Feature Reduction}\label{section:simple_feature_selection}
We first look at ``simple'' feature reduction: we manually select features based on how they are expressed by the applications in our data set. We formulate these simple feature reduction strategies through \emph{feature restrictions} that we impose on the feature set that we use for training our neural networks.

The first feature restriction is that we train only on features that are only expressed in the \texttt{AnroidManifest.xml} of the applications' APKs. Recall that, in the adversarial crafting algorithm, we decided to only add features that appear in the manifest to make sure that we do not interfere with any functionality of the application in question. By restricting the training of our neural networks to manifest features only, we focus the classifier on the features that we actually change. We denote this feature restriction with \monly. By applying \monly, we are left with $233727$ of the original $545333$ features that we use for training.

As the second feature restriction, we only consider features that do not appear in the $r$ largest feature classes. We illustrate the cardinalities of our $8$ feature classes in Table~\ref{table:feature_types}. A large part of the features (over $50\%$) consist 
of URLs that are mostly unique to each application, and would therefore not help with separating benign from malicious applications. Other features that appear in the largest feature classes show similar behavior. With this feature restriction, which we call \onlys, we try to define a simple way to filter out all such features. We instantiate \onlys for $r=1$ and $2$, which leaves us with $234845$ and $49116$ features, respectively, after applying the feature restriction.

The last feature restriction that we consider is a variant of \onlys. With this restriction, we only consider features that appear in at least $r$ applications. We thereby directly filter out features that do not consistently appear across representatives of one target class, and will therefore not be used to detect an application that belongs in them. We call this feature restriction \onlyf. 

In effect, \onlyf tries to achieve a similar goal as \onlys, however the set of restricted features is determined differently. This is also reflected in the number of features that remain after the applying the restriction: we instantiate \onlyf for $r\in\{1,2,3,4\}$, which leaves us with $177438$, $95871$, $60052$ and $44942$ features respectively.
\begin{figure}[t]
\includegraphics[width=\linewidth]{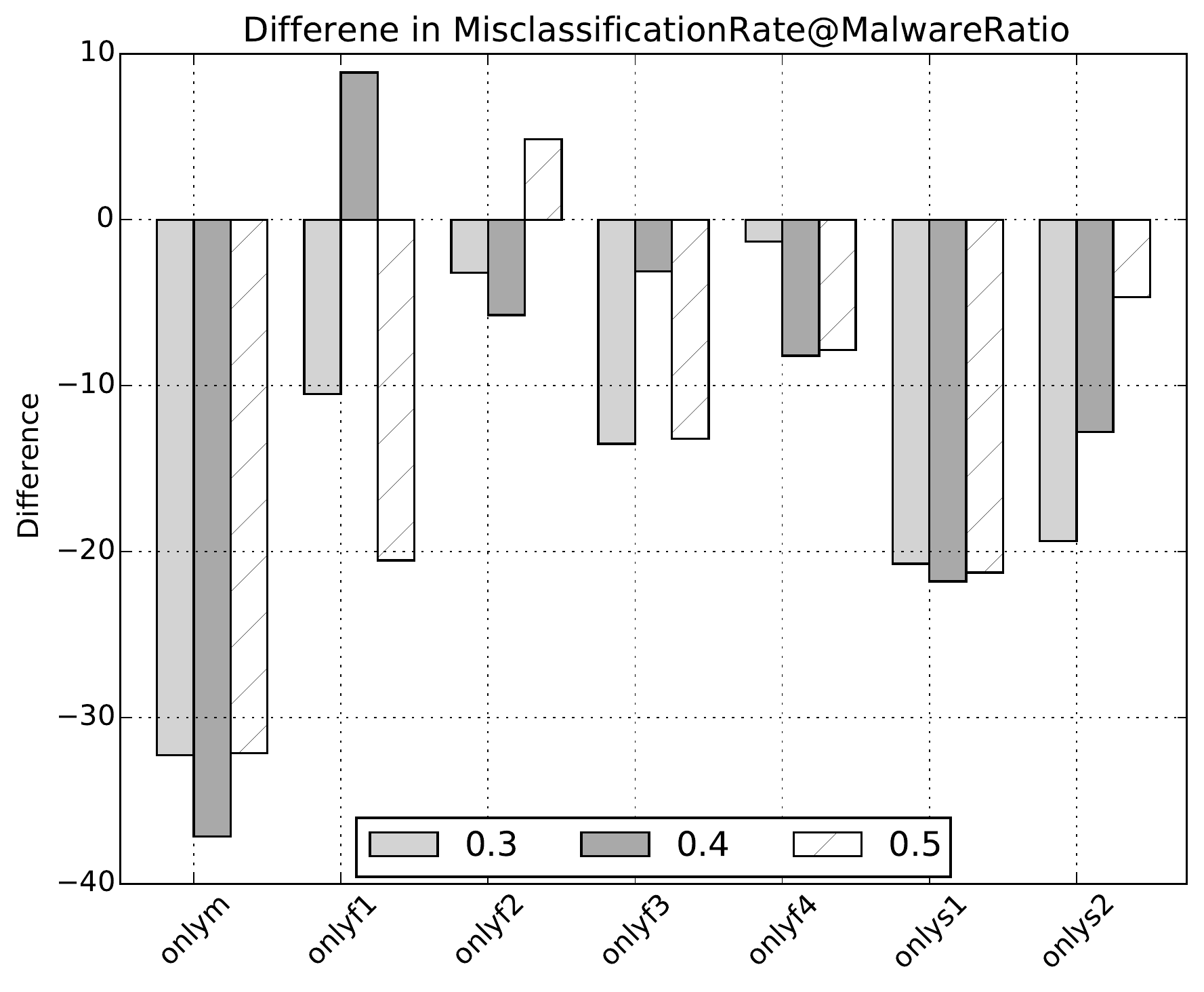}
\caption{Difference in misclassification rates for the different simple feature reduction methods, with regular networks as baseline. Negative values indicate a higher misclassification rate, 
and therefore a more susceptible network.}\label{fig:simple_feature_reduction}
\end{figure}
\subsubsection{Evaluation}
We use all three simple feature reduction methods to train new neural networks, each with 2 hidden layers and 200 neurons on each layer. We then run our adversarial crafting algorithm on each of these networks to determine how frequently they are mislead by adversarial crafting. Figure~\ref{fig:simple_feature_reduction} shows the difference between the original misclassification rates we achieves on the regular networks and the misclassification rates we achieve for each feature reduction method. This difference is evaluated at the different malware ratios we used for training.

As we can see, the simple feature reduction generally weakens the neural network against adversarial crafting. In contrast to the about 
$62\%$ misclassification rate we achieve on the regular neural network, the misclassification rate goes up to, in the worst case, $99\%$ for the \monly feature restriction. Most other neural networks trained under feature restrictions also show an increase of the misclassification rate. While there are two exceptions, for instance under the restriction \onlyf with $r=1$ where we reduce the misclassification rate by around $8\%$, these cannot be observed consistently across architectures. We therefore leave them as exceptions to the rule.

Overall, we can conclude that simple feature reduction is not a suitable mechanism to harden your neural network against adversarial crafting in the domain of malware detection. 
\subsection{Feature Reduction via Mutual Information}\label{section:mutual_information}
Next, take a look at feature reduction using mutual information, instead of taking the naive approach we described in the previous section. We first briefly introduce the notion of mutual information, 
before we then investigate its influence on our networks' resistance against adversarial crafting.

The mutual information $I(X;Y)$ of two random variables $X$ and $Y$ is given by
 \begin{equation}
 I(X;Y)=\sum_{X,Y}p(x,y)\log(\frac{p(x,y)}{p(x)p(y)}).
 \end{equation} 
Intuitively, the mutual information $I(X;Y)$ of two random variables $X$ and $Y$ quantifies the statistical dependence between $X$ and $Y$. We use this concept to identify features in our data set that carry a large amount of information with regard to our target classes. 
To select features using mutual information, we compute it for each feature/target class pair, i.e. we compute the mutual information $I(X;Y)$, where $X$ is the probability of a feature and $Y$ the probability of a target class. We then order all features by their mutual information with the target classes and we pick the $n$ highest of them. If we encounter a tie, we include all features with that same mutual information (and thus obtain a number of features slightly bigger than $n$). We finally use the set of features determined by this process to train new classifiers and evaluate their resistance against adversarial crafting.

Consider that there is a second approach, which we also investigate. Given that a feature has a high value in mutual information, we also assume that its influence on the classification is high. We will thus train a classifier only on feature sets that contain many but equally important features and observe the misclassification rates.To do so, we inverse the ranking from the first experiment and consider with $n$ the $n$ lowest values of mutual information and their corresponding features. 

There are few features with a high value of mutual information and many with a low value. As a consequence, the second approach will yield much bigger datasets to train on then the first.

\begin{figure}
\includegraphics[width=\linewidth]{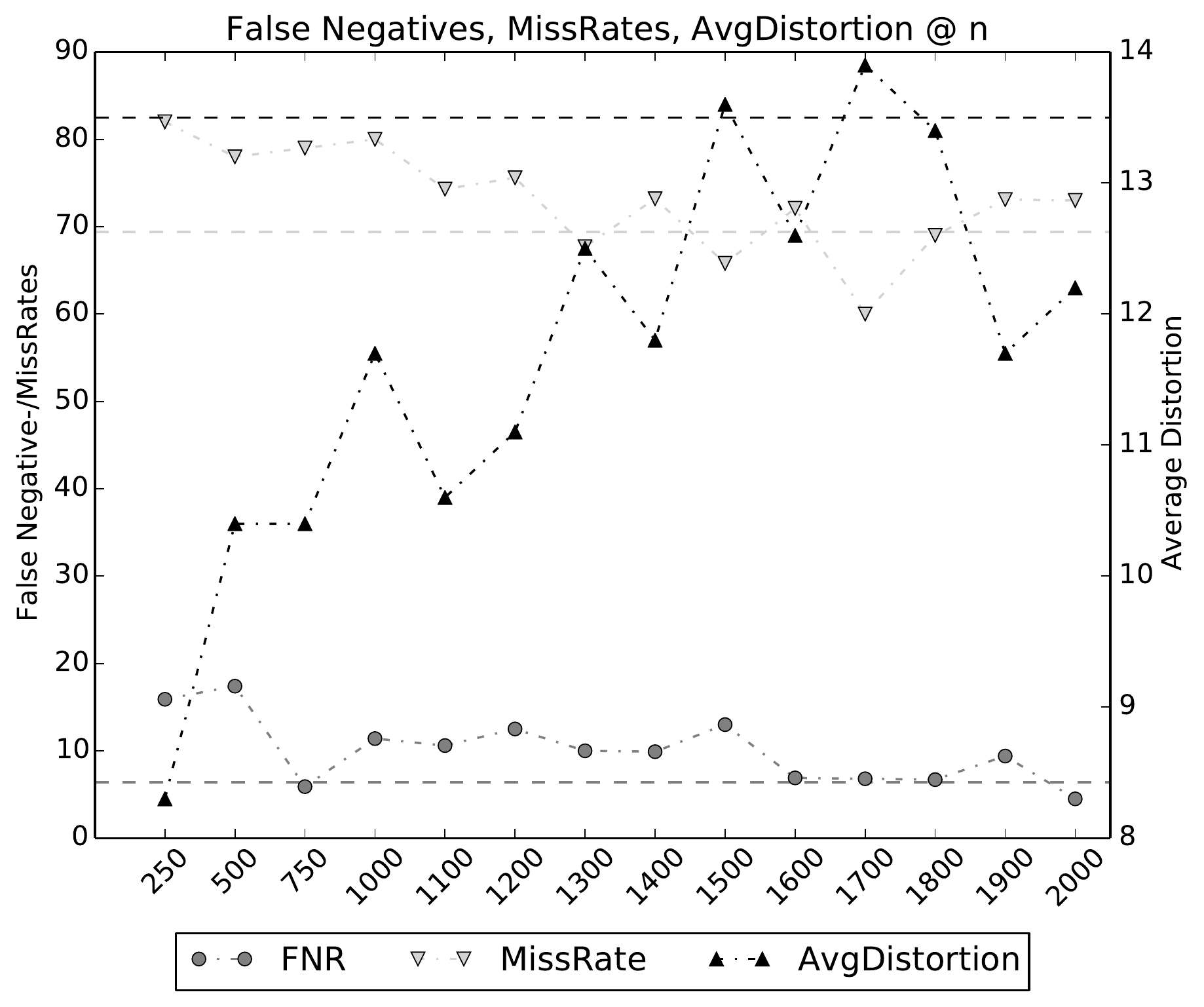}
\caption{Success of adversarial crafting against classifiers trained under feature reduction using mutual information. Given are false negative rates, misclassification rates and average distortion on each of these networks at mutual information level $n$. All networks have shape [200,200] and are trained with a malware ratio of 0.5. The constant values are the corresponding rates of the original model.}\label{fig:MI}
\end{figure}
\subsubsection{Evaluation first Approach}
In this approach we rank after the importance of features given MI. We trained several new neural networks for varying values of $n$, ranging from $n=50$ to $n=2000$. Each of these networks use the standard architecture (i.e. 2 hidden layers with 200 neurons each) and are trained with a $50\%$ malware ratio during training. We then applied the adversarial crafting algorithm described in Section~\ref{section:crafting_samples} to determine their susceptibility to the attack. 

Our evaluation results are presented in Figure~\ref{fig:MI}. It shows the development of false negative rates (at accuracy > $93\%$), misclassification rates and average required distortion for an increasing amount of features selected through the mutual information method described above. As a point of comparison, we provide the performance values of the regular neural network with feature reduction as horizontal lines (i.e. $7\%$ false negatives, $62\%$ misclassification rate and an average distortion of around $15$).   

For very small $n<250$, we experience very poor classification performance and therefore ignore those cases in our evaluation. For the other cases, performance is in general worse than in the regular case. For $n\leq 1600$, we only achieve about $7\%$ false negative rates in the classification, and the adversarial crafting algorithm is successful in at least $70\%$ of the cases, requiring a smaller average distortion of $12.5$.

For $1500<n < 1900$, we observe a peak where our newly trained networks achieve nearly the same performance as our regular networks: the false negative rate is around $7\%$ and the misclassification rate also goes down to about $60\%$ for $n=1700$. The required average distortion, however remains low at at most $14$. The misclassification rates differ however strongly for the similar models.
\subsubsection{Evaluation second Approach}
\begin{figure}
\includegraphics[width=\linewidth]{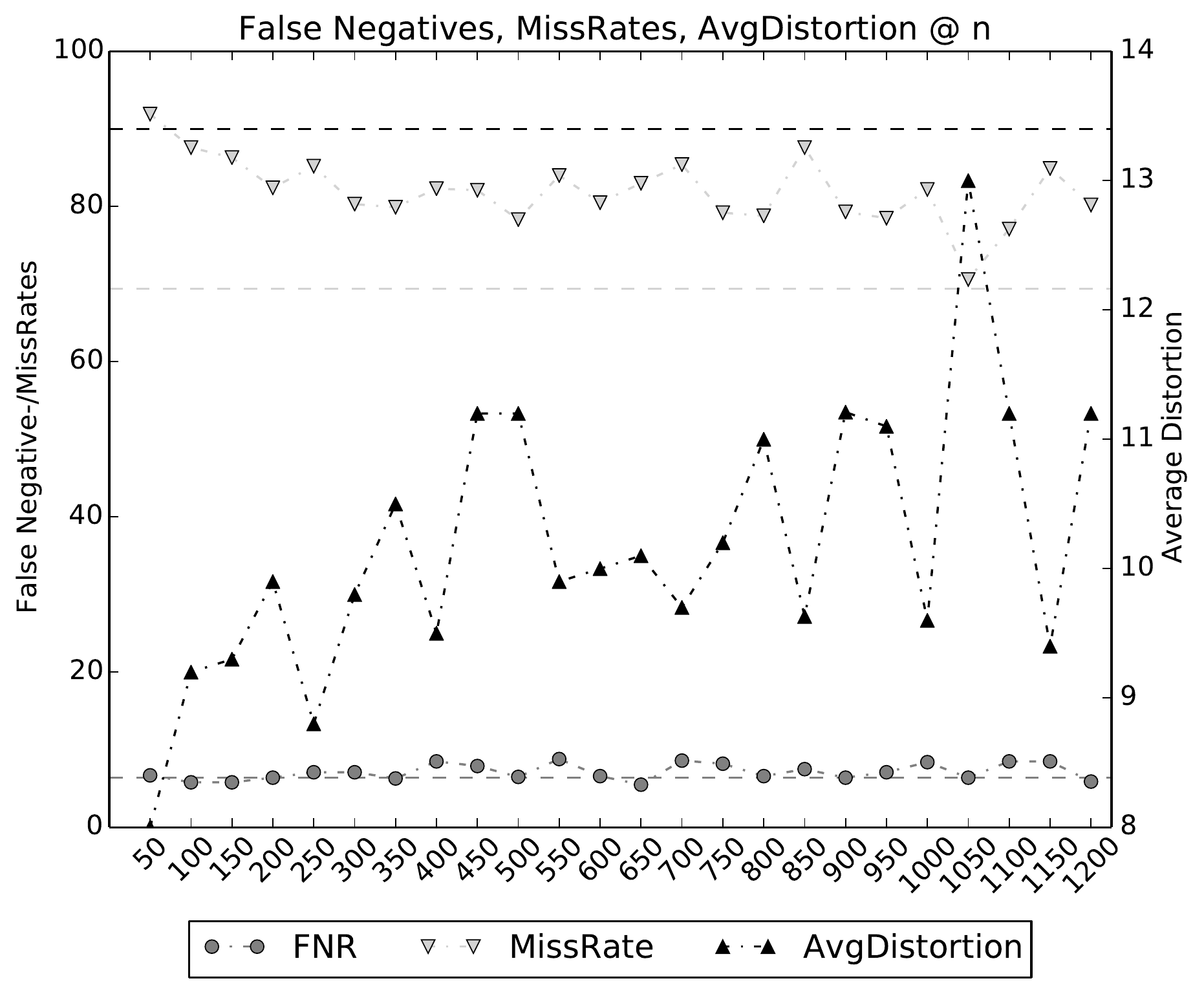}
\caption{Success of adversarial crafting against classifiers trained under feature reduction using mutual information in reversed order. Given are false negative rates, misclassification rates and average distortion on each of these networks at mutual information level $n$. All networks have shape [200,200] and are trained with a malware ratio of 0.5. The constant values are the corresponding rates of the original model.}\label{fig:revMI}
\end{figure}

We again trained several networks on different values of $n$, where $n$ however now starts with the tied, low ranked features. For very low $n$, we observe that the network is very vulnerable to adversarial samples, having a misclassification rate of over $90\%$. 

The false negative and accuracy rates are, however, in general comparable with the original model trained on the full data. The general misclassification rate decreases with more features and varies between $70.1$\% for $n=1050$ and $87.6$\% for $n=800$, average is $82$\%. This is much worse than the original model.

\subsubsection{Discussion}
As in the previous section, our findings indicate that using less features does not necessarily make the adversarial crafting harder. It seems that few numbers features lead to easy perturbations, since each single feature has a much higher impact for classifying an application. By using feature reduction based on mutual information, we thus make our network even more susceptible against adversarial crafting.
 
In the case of selecting the most important features, however, there are some cases where the trained model is less prone to adversarial samples. Hence, we do not want to exclude the possibility of improving resistance against adversarial crafting using feature reduction: a more fine-grained analysis could lead to positive results that we leave for future work.

 

\subsection{Distillation}\label{section:distillation}
We next take a look at \emph{distillation}. While distillation was originally proposed 
by Hinton et al.~\cite{2015arXiv150302531H} as as a way to transfer knowledge from large neural networks to a smaller ones, 
Papernot at al.~\cite{papernot2016distillation} recently proposed using it as a defensive mechanism against adversarial crafting as well.

The basic idea behind distillation is the following: assume that we already have a neural network $\F$ that classifies a training data 
set $\samples$ into the target classes $\classes$ and produces as output a probability distribution over $\classes$ (e.g. by using a final softmax layer we already introduced in Section~\ref{section:training_classifier}). Further, assume that we 
want to train a second neural network $\F'$ on the same data set $\samples$ that achieves the same (or even better) performance. 
Now, instead of using the target class labels that come with the data set, we use the output $\F(\X)$ 
of the network $\F$ as the label of $\X\in\samples$ to train $\F'$. The output produced by $\F$ will 
not assign a unique label $\Y\in\classes$ to $\X$, but instead a probability distribution over $\classes$. 
The new labels therefore contain more information about the membership of $\X$ to the different target classes, 
compared to a simple label that just chooses the most likely class.

Papernot et al.~\cite{papernot2016distillation} motivate the use of distillation as a defensive mechanism through its capability to improve the second networks generalization performance (i.e. classification performance on samples outside the training data set). Thus, instead of  using distillation to train a second smaller network like was proposed in~\cite{2015arXiv150302531H}, they use the output of the first neural network $\F$ to train a second neural network $\F'$ with exactly the same architecture.   


An important detail in the distillation process is the slight modification of the final softmax layer (cf. Equation~\ref{eq:softmax}) in the original network $\F$: instead of the regular softmax normalization, we use
\begin{equation}
\F_i(X) = (\frac{e^{z_i(x)/T}}{\sum^{\lvert \classes \rvert}_{l=1}e^{z_l(x)/T}}),
\end{equation}  
where $T$ is a distillation parameter called \emph{temperature}. For $T=1$, we obtain the regular softmax normalization that we already used for the training in Section~\ref{section:training_classifier}. If $T$ is large, the output probabilities approach a more uniform distribution, whereas for small $T$, the output of $\F$ will be less so. To achieve a good distillation result, we use the output of the original network $\F$ produced at a high temperature $T$ and use it as class labels to train the new network $\F'$.
The overall procedure for hardening our classifier against adversarial crafting can thus be summarized in the following three steps.
\begin{enumerate}
\item Given the original classifier $\F$ and the samples $\samples$, construct a new training data set $D=\{(\X,\F(X)) ~\vert~ \X \in \samples\}$ that is labeled with $\F$'s output at high temperature.
\item Construct a new neural network $\F'$ with the same architecture as $\F$.
\item Train $\F'$ on $D$.
\end{enumerate}
Note that both step two and step three are performed under the same high temperature $T$ to achieve a good distillation performance. 

We next apply the above procedure on our originally trained classifiers and examine the impact of distillation as a defensive mechanism against adversarial crafting in the domain of malware detection.
\subsubsection{Evaluation}
\begin{figure}
\includegraphics[width=\linewidth]{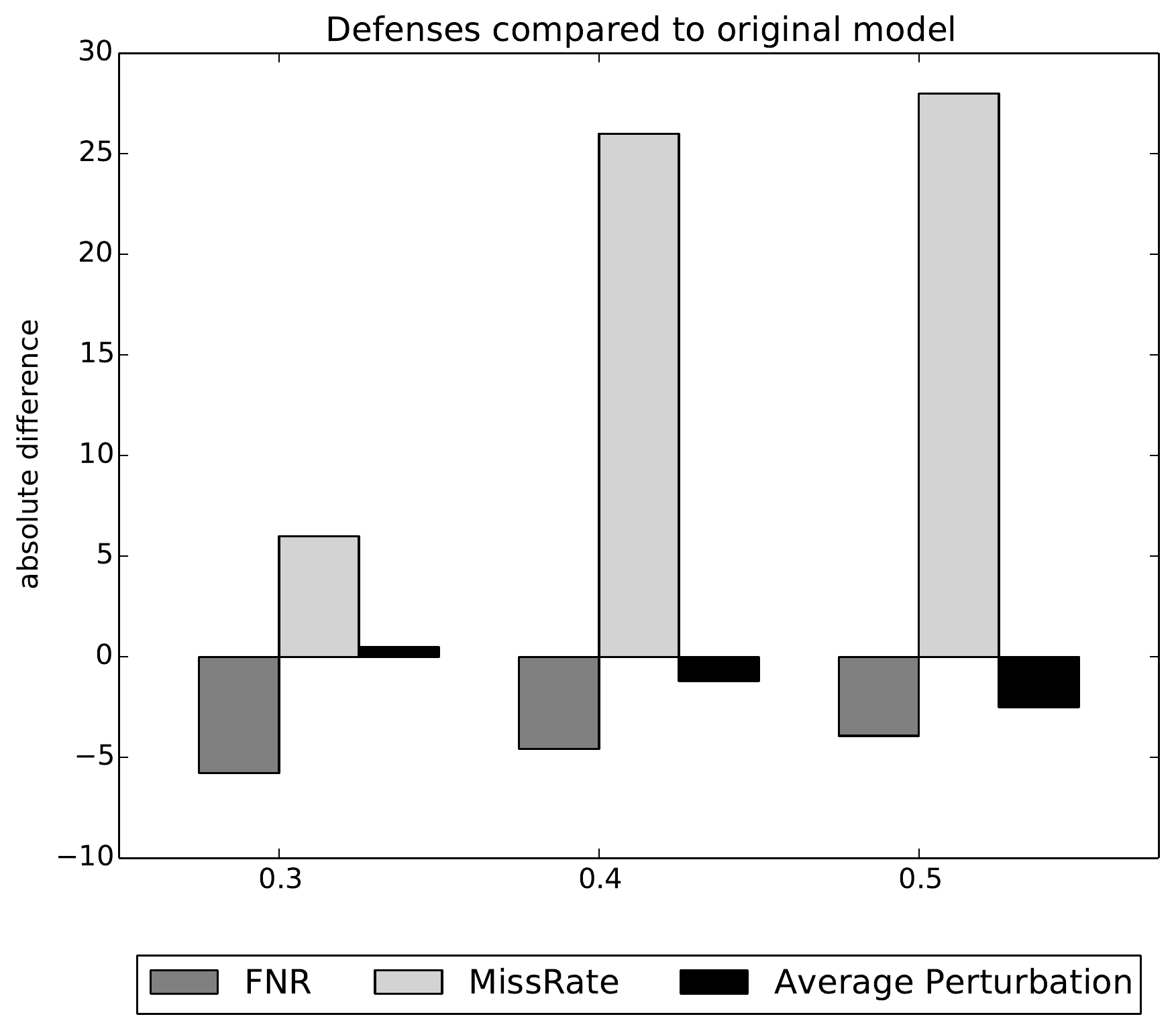}
\caption{False negative rates, misclassification rates and average required distortions after applying distillation to the networks networks with 200 neurons on 2 layers each. Given are the differences of each measure to the original network's performance. A positive value is thus an improvement for FNR and misclassification rate, value for average distortion should be negative.}\label{fig:dist}
\end{figure}
Figure~\ref{fig:dist} shows the effects of distillation on misclassification compared to the original models. We observe a general, strong increase of the false negative rate, and a slight increase in the false positive rate. For ratio $0.5$, it raises from $4$ to $6.4$, whereas for $0.3$, it is equivalent. Due to the large size of the benign class, the accuracy only ranges in between $93$-$95$\%.

On the other hand, we observe that the misclassification rate drops significantly, in some cases to $38.5$\% for ratio $0.4$. The difference in the average number of perturbed features, however, is rather small. The number of perturbed features is $14$ for ratio $0.3$ to $16$ for the other two.

Using distillation, we can strengthen the neural network against adversarial samples. However, the misclassification rates are still around $40\%$ and thus rather high. Additionally, we pay this robustness with a less good classifier. The effect is further not as strong as on computer vision data. Papernot et al.~\cite{papernot2016distillation} reported rates around $5\%$ after distillation for images. Further research should investigate whether distillation can be adapted to malware data or discrete, sparse and unbalanced data in general. Also the question remains which properties (or combinations thereof) influence the obtained improvement in terms of misclassification rates.

\subsection{Re-Training}\label{section:re-training}
As the last countermeasure, we try re-training our classifier with adversarially crafted samples. This method was originally proposed by
Szegedy at al.~\cite{szegedy2013intriguing} and involves the following steps:
\begin{enumerate}
\item Train the classifier $\F$ on original data set $D = B \cup M$, where $B$ is the set of benign, and $M$ the set of malicious applications
\item Craft adversarial samples $A$ for $\F$ using the forward gradient method described in Section~\ref{section:crafting_samples}
\item Iterate additional training epochs on $\F$ with the adversarial samples from the last step as additional, malicious samples.
\end{enumerate}
By re-training, we aim at improving the generalization performance of $\F$, i.e. improve the classification performance of $\F$ on samples outside our training set. A good generalization performance generally makes a classifier less sensitive to small perturbations, and therefore also more resilient to adversarial crafting.

\subsubsection{Evaluation}
We applied the above mechanism to the regular networks with 200 neurons on 2 layers each that we have been consider throughout this section. We continued their training on $n_1 = 20$, $n_2=100$ and $n_3=250$ additional, adversarially crafted malware samples. We combined the adversarial samples to create training batches by mixing them with benign samples at each network's malware ratio. We then trained the network for one more epoch on one training batch and re-evaluated their susceptibility against adversarial crafting.

\begin{figure}
\includegraphics[width=\linewidth]{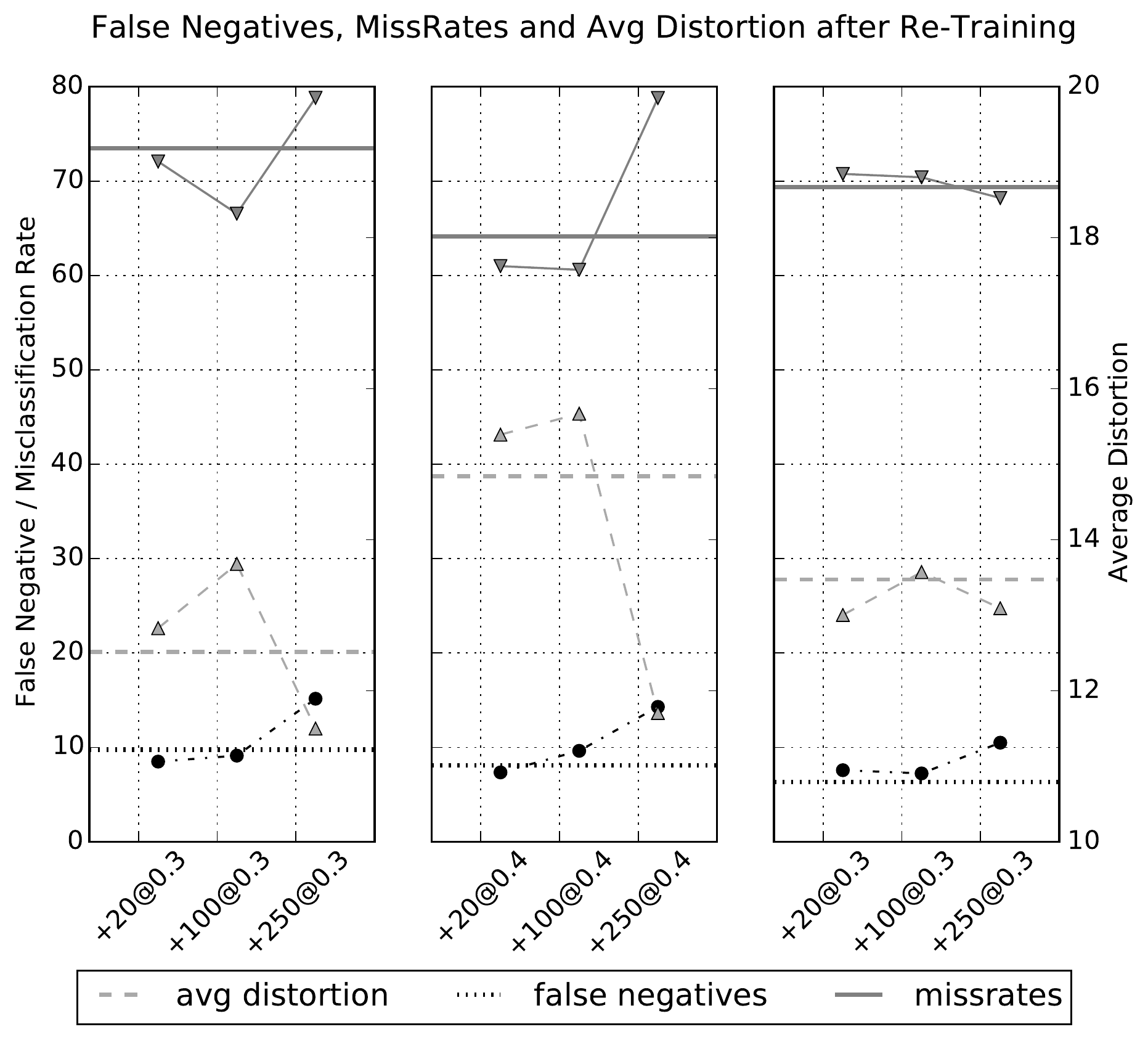}
\caption{Misclassification rates, false negative rates and average required distortion achieved on re-trained networks. Regular network's performance is given as baseline, indicated by horizontal lines.}\label{fig:misclassification_re-training}
\end{figure}

Figure~\ref{fig:misclassification_re-training} illustrates the performance (in false negative rate) of the re-trained networks and the misclassification performance 
Algorithm~\ref{sampling_algorithm} achieved on them (in misclassification rate and average required distortion). We grouped networks by their malware ratio during training and give as 
comparison the performance values of the original networks before re-training. 

For the network trained with malware ratio $0.3$ and $0.4$, we can observe a reduction of the misclassification rate, and an increase of the required average distortion for $n_1$ and $n_2$ additional training samples. For instance, we achieve a misclassification rate of $67\%$ for the network trained with $100$ additional samples at $0.3$ malware ratio, down from $73\%$ for the original network. A further increase of the adversarial training samples used for re-training, however, causes the misclassification rate to increase again, reaching up to $79\%$ for both malware ratios.

For the last networks, trained with malware ratio $0.5$, the misclassification rate only decreases if we use $250$ adversarial training samples. Here, we reach $68\%$ 
misclassification rate, down from $69\%$ for the original network. For fewer amount of adversarial samples for re-training, the misclassification rate remains very similar 
to the original case. It seems that the network trained with $0.5$ malware ratio is fitting very close to the malware samples it was trained on, and therefore requires more 
adversarial samples to generalize and improve its resistance against adversarial crafting.  

Overall, we can conclude that simple re-training on adversarially crafted malware samples does improve the neural network's resistance against adversarial crafting. The 
number of adversarial samples required to improve the resistance depend heavily on the training parameters we chose for training the original networks. However, choosing too 
many may also further degrade the network's resistance against adversarial crafting. 


\subsubsection{Discussion}
In our evaluation above, we only considered one iteration of adversarial re-training. Ideally, the above method is continuously applied, by computing new adversarial 
samples for each newly generated network $\F'$.
These samples are then fed back to re-train $\F'$ and generate a new network more resistant against adversarial crafting.  

Goodfellow et al.~\cite{goodfellow2015explaining} propose an alternative approach to the general idea of adversarial
re-training: instead of training on actually crafted adversarial samples, they formulate an adversarial loss function 
that incorporates the possibility of adversarial crafting through perturbations in direction of the network's gradient. 
This allows them to continuously incorporate adversarially crafted samples during training with an infinite supply thereof. 
We think that this approach should further be investigated also in security critical domains, and consider this a promising direction for future work.

\subsection{Summary of Results}\label{section:defense_discussion}
We considered four potential defensive mechanisms and evaluated their impact on a neural networks susceptibility against adversarial crafting. Feature reduction, both simple as well through 
mutual information, usually make the neural network weaker against adversarial crafting. Having less features of greater importance makes it easier to craft adversarial samples. This is caused by the larger impact each feature has on the output distribution of the network. At this point we cannot recommend
feature reduction as a defensive mechanism. Future work will have to identify, potentially more involved, feature reduction methods that actually 
increase a network's resistance against adversarial crafting. 

We also investigated distillation and re-training, which were both originally proposed as defensive mechanism against adversarial crafting in the computer vision domain. Distillation does have a positive effect, but does not perform as well as in the computer vision domain. The reasons for this have to be investigated in future work. Simple re-training achieved consistent reduction of misclassification rates across different networks. However, choosing the right amount of adversarial training samples has a significant impact on this reduction. Iteratively applying re-training to a network might further improve the network's resistance.

\section{Related Work\ifdraft (1 Column)\fi}\label{section:related_work}
The following discussion of related work complements the references included in Section~\ref{section:background}.

Security of Machine Learning is an active research area. Barreno et al.~\cite{DBLP:journals/ml/BarrenoNJT10} give a broad overview of attacks against machine learning systems. They discriminate between exploratory attacks at test time or causative attacks that influence the training data to obtain the desired result. Adversarial samples, as used here, are employed at test time. Previous work showed that adversarial samples can be constructed for different algorithms and also generalize between machine learning techniques in many cases \cite{goodfellow2015explaining,papernot2016practical,szegedy2013intriguing}.

Gu et al.~\cite{DBLP:journals/corr/GuR14} claim that adversarial samples are mainly a product of the way feed-forward neural networks are trained and optimized. As a solution they propose deep contractive networks, which are however harder to optimize. These networks include a layer wise penalty which furthermore limits their capacity.

Goodfellow at al.~\cite{goodfellow2015explaining} proposed a linear explanation to the existence of adversarial samples. In their intuition, adversarial samples are not due to the overall non-linearity of neural networks, but to the linearity of their underlying components. Consequently, they show that adversarial samples generalize to other linear models, such as for example logistic regression. In contrast to those models, neural networks are however able to be hardened against adversarial samples.

\section{Conclusion and Future Work}\label{section:conclusion}
In this paper, we investigated the viability of adversarial crafting against neural networks in domains different from computer vision. Through our evaluation on the DREBIN data set, we were able to show that adversarial crafting is indeed a real threat in security critical domains such as malware detection as well: we achieved misclassification rates of up to $80\%$ against neural network based classifiers that achieve  classification performance on par with state of the art classifiers from the literature.

As a second contribution, we examined four potential defensive mechanisms for hardening our neural networks against adversarial crafting. 
Our evaluations of these mechanisms showed the following: first, feature reduction, a popular method to reduce input complexity and simplify the training of the classifier, generally makes the neural network weaker against adversarial crafting. Second, distillation does improve misclassification rates, but does not decrease them as strongly as observed in computer vision settings.
Third, re-training on adversarially crafted samples achieves consistent reduction of misclassification rates across architectures.

Future work should more carefully examine the defensive mechanisms that we identified as potentially helpful For example in addition to distillation, Goodfellow et al.'s~\cite{goodfellow2015explaining} idea of using an adversarial loss functions 
during training should be more carefully examined in security relevant domains.

The applicability of  adversarial crafting attacks to additional domains should also be studied. It could direct researchers towards more effective defensive mechanisms.

\let\OLDthebibliography\thebibliography
\renewcommand\thebibliography[1]{
  \OLDthebibliography{#1}
  \setlength{\parskip}{1.3ex}
  \setlength{\itemsep}{0.3ex}
}

\bibliographystyle{latex8}
\bibliography{lit}

\end{document}